# Progressive Prediction of Turbulence Using Wave-Front Sensor Data in Adaptive Optics Using Data Mining

Akondi Vyas[1], M B Roopashree[2], B Raghavendra Prasad[3]

[1]Research Scholar, Indian Institute of Astrophysics, Bangalore-34
[2]Research Scholar, Indian Institute of Astrophysics, Bangalore-34
[3]Professor, Indian Institute of Astrophysics, Bangalore-34
vyas@iiap.res.in, roopashree@iiap.res.in, brp@iiap.res.in.

## ABSTRACT

*Nullifying the servo bandwidth errors improves the strehl ratio by a substantial quantity in adaptive optics systems. An effective method for predicting atmospheric turbulence to reduce servo bandwidth errors in real time closed loop correction systems is presented using data mining. Temporally evolving phase screens are simulated using Kolmogorov statistics and used for data analysis. A data cube is formed out of the simulated time series. Partial data is used to predict the subsequent phase screens using the progressive prediction method. The evolution of the phase amplitude at individual pixels is segmented by implementing the segmentation algorithms and prediction was made using linear as well as non linear regression. In this method, the data cube is augmented with the incoming wave-front sensor data and the newly formed data cube is used for further prediction. The statistics of the prediction method is studied under different experimental parameters like segment size, decorrelation timescales of turbulence and segmentation procedure. On an average, 6% improvement is seen in the wave-front correction after progressive prediction using data mining.*

## 1. INTRODUCTION

Data mining is a useful technology in predicting future trends and behaviors of temporal data patterns. Predicting the future trends in business is one of the most commonly used applications of temporal data mining strategies [1]. The realization of the importance of data mining in astronomical community has led to many useful results [2-4]. Adaptive optics has become an indispensable technology used in large telescopes to improve the image quality, degraded due to the presence of atmospheric disturbances [5]. In an astronomical adaptive optics system, the temporally induced turbulence has to be corrected in real time with the help of a correcting element, generally a deformable mirror. The information of the shape of the wave-front is detected by the wave-front sensor. The control algorithm then calculates the command values to be addressed to the actuators of the deformable mirror that conjugates the ill effect of turbulence. Prediction of turbulence gained significance in the case of adaptive optics to remove the servo lag error [6].

In this paper, the atmospheric turbulence is modeled such a way that it follows Kolmogorov spatial statistics. A time series of evolving turbulence is simulated using the wind model as described in the next section. The series of phase screens are then stacked to form a large data cube. A small portion of the large data cube is chosen for prediction. The evolution of phase at individual pixels is segmented using the bottom up algorithm in the smaller data cube. Regression is performed to estimate the value of phase of the next subsequent phase screen at each of the pixels. A progressive prediction model is used to estimate future turbulence screens. The performance of the prediction is limited by the segmentation parameters, the polynomial fitting, the servo lag timescales and finally the data statistical errors. The computational results of the effectiveness of prediction are presented.

## 2. MODELLING ATMOSPHERIC TURBULENCE

Any phase function f(x, y) can be written as a linear combination of the basis set of Zernike polynomials, $Z_i(x, y)$.

$$f(x,y) = \sum_{i=1}^{\infty} a_i Z_i(x, y) \qquad (1)$$

where, $a_i$ represents the Zernike moments. Atmospheric turbulence can be simulated using the Kolmogorov theory of turbulence [7]. Noll established a relationship between the spatial statistics of Kolmogorov and the correlation statistics of the Zernike moments. In order to simulate phase screens that satisfy the statistical theory of Kolmogorov, the Zernike moments must be determined by the relationship [8],

$$<a_i a_j> = 0.0072 \left(\frac{D}{r_0}\right)^{5/3} (-1)^{(n_i+n_j-2m_i)/2} \pi^{8/3}$$
$$\delta_{m_i m_j} \Gamma[(n_i+n_j-5/3)/2]\{(n_i+1)(n_j+1)\}^{1/2}$$
$$\{\Gamma[(n_i-n_j+17/3)/2]\,\Gamma[(n_j-n_i+17/3)/2]$$
$$\Gamma[(n_i+n_j+23/3)/2]\}^{-1}\Gamma(14/3)$$

(2)

D is the diameter of the telescope; $r_0$ is the atmospheric seeing parameter; $n$ and $m$ are the radial and azimuthal indices of Zernike polynomials. Calculation of the turbulence function through the computation of Zernike moments using eq. (2) gives frozen in turbulence for a given atmospheric seeing condition. The time-evolving turbulence wave-fronts cannot be obtained by just changing the Zernike coefficients, because the produced Zernike coefficients lead to temporally uncorrelated wave-fronts. To simulate dynamic turbulence, we can use the simple properties of the propagation of wind in the atmosphere as suggested earlier [9].

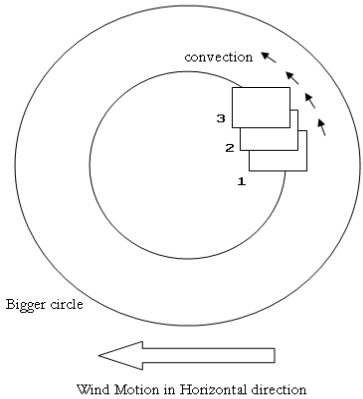

Fig. 1. Modeling the temporal evolution of atmospheric turbulence

The propagation properties of wind that can be considered are the motion of wind in the horizontal direction and the convection of wind as shown in Figure 1. As a first step we simulate a large wave-front. A smaller part of this large wave-front is then read. In the third step, another part of the large wave-front which is a few pixels away, defined by the rotational and translational velocities at that time is read. Third step is repeated to simulate a series of wave-fronts replicating atmospheric turbulence. The series of phase screens are represented as a time series A = $\{A_j\}$, j takes integral values from 1 to N, where N is the number of phase screens read from the large phase screen. For simplicity, the velocities are assumed to be constant over time. A few sample turbulence phase screens evolving in time are shown in Figure 2.

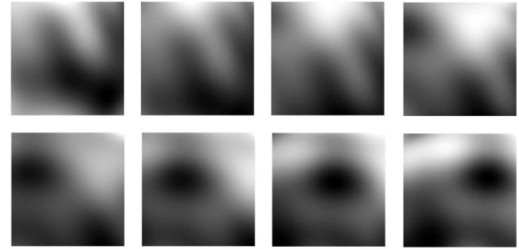

Fig. 2. Series of evolving turbulence phase screens

The correlation between the first phase screen and the subsequent phase screens varies as shown in the graph in Figure 3. Since any two phase screens randomly chosen also can be correlated by 25%, we can conveniently assume that the phase screens correlated by less than 25% are decorrelated.

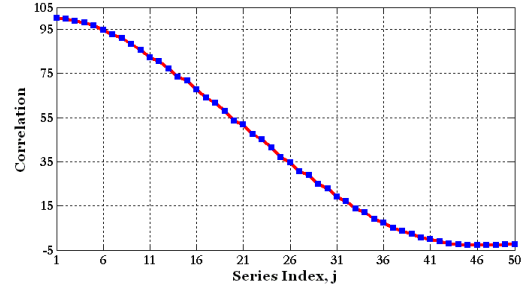

Fig. 3. Correlation of subsequent phase screens in the time series

## 3. FORMATION OF DATA CUBES

The formed phase screens are pixelated two dimensional arrays containing gray scale values. To predict the subsequent phase screens, it is important to study the variation of the gray scale value of a particular pixel in time. In order to club the pixel gray scale data, data cubes can be formed by stacking the phase screens in the same order as they appear in the time series as shown in Figure 4. The number of phase screens used to form the data cube is called the length of the data cube.

The data cube formed in this fashion can be used for further analysis. Data sets are formed by collecting the temporal evolution of the phase amplitude corresponding to individual pixels. There are as many data sets as the number of pixels. Segmentation algorithms are applied on these individual data sets and interpolation techniques are used to predict the future expectation value corresponding to individual pixels. To overcome space and speed problems due to over stacking, the older wave-front sensor data is removed while adding new data and the capacity of the data cube at any point of time is maintained a constant. This is the reason why it is called the progressive prediction method.

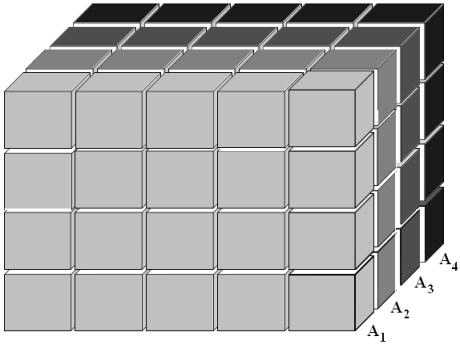

Fig. 4. Data cube formed out of the time series database

## 4. SEGMENTATION METHODOLOGY

Piecewise linear segmentation is one of the commonly used time series representation [10]. In the piecewise linear representation, the time series data is segmented into smaller pieces and each of them is fitted with a straight line. There are many algorithms to segment the data like sliding windows, top-down and bottom-up algorithms. In the sliding windows algorithm, a particular segment is grown until it reaches an error threshold. Beyond the error threshold, a new segment is begun. As the name suggests, in the top-down algorithm, the time series is broken recursively until it goes below the threshold error. The bottom-up algorithm is a compliment of top-down algorithm wherein segmenting is done starting from the smallest possible interval and adjacent segments are merged until the error is just below the error threshold. It is experimentally verified that top-down and bottom-up algorithms have an edge over sliding windows in terms of their performance [10].

Above mentioned segmentation algorithms are used for segmentation process in the case of simulated Kolmogorov turbulence time series. The various parameters in segmentation that affect the performance of prediction algorithm include the segment size, the spatial and temporal scales of variations in turbulence and the error threshold. Some other parameters like the number of pixels and the turbulence simulation model are not considered in our study which might have considerable significance in terms of prediction accuracy.

Segmentation is generally associated with piecewise linear representation. Linear interpolation as well as linear regression is performed on individual segments after using one of the methodologies for segmentation. Using nonlinear regression on segments gives absurd results when the length of individual segments is small. Higher order regression on larger segments will reduce the fitting error and effectively minimizes the time by reducing the number of segments.

## 5. PROGRESSIVE PREDICTION

Prediction of turbulence in adaptive optics correction of the blurring of images due to atmospheric turbulence and dynamic processes in the human eye is helpful in reducing the servo lag error. This error is introduced by the extreme bounds imposed by the system servo bandwidth that is defined as the temporal frequency at which the real time closed loop corrections take place [11].

The upper bound on the bandwidth is limited by two factors; one is the finite response times of the sensor, corrector and the control system and the second is the minimum exposure time required to lessen the wave-front sensor data errors. The second upper bound is applicable in the case of atmospheric turbulence case where the exposure timescales are nearly 5 milliseconds when imaging a natural guide star and 1 millisecond in the case of laser guide star. The lower bound is the decorrelation time, defined as the time scale over which the wave-fronts become decorrelated. The decorrelation time, $\tau$ is defined as

$$\tau = 0.31 r_0 / v_w \qquad (3)$$

where $r_0$ is the Fried parameter and $v_w$ is the wind velocity [11]. For the Hanle site, the decorrelation timescale, $t_D$ is nearly 18 milliseconds [12]. Within $t_d$, the wave-fronts remain correlated. Since the exposure time is finite (nearly 1 millisecond even for a Laser Guide Star), and there will be a lag due to the reconstruction of the wave-front from sensor data and generation of actuator command values to be addressed to the deformable mirror, the wave-front that is sensed and the wave-front that is being corrected are uncorrelated by a certain amount. This incorrect compensation leads to imperfect correction of the atmospheric aberrations.

A possible solution for this problem is to progressively predict the wave-front that might be arriving at a servo lag time, '$t_L$' later by using the available wave-front sensor information. According to our knowledge, data mining is explored for the first time in the case of adaptive optics applications. This can be done by following the three steps shown below:
1) Segmentation of the individual temporal pixel data sets that are picked up from the data cube.
2) Predicting the pixel value of the next phase screen using the regression formula for the last segment in individual pixel data sets.
3) Forming the phase screen by combining the predicted pixel values.

The prediction process is progressive in nature and each time the wave-front sensor data is obtained; it is concatenated at the end of the data cube. The size of the data cube is retained by removing the same number of phase screens from the other end.

## 6. EXPERIMENTS AND RESULTS

Monte Carlo simulations were implemented by following the three steps suggested in the last section in the case of the time series of the simulated Kolmogorov phase screens. As hinted earlier, the degree of accuracy of prediction depends on the segment size or the error threshold chosen for segmentation, the order of polynomial fitting, the servo lag timescales and decorrelation time. Data cubes of length 350 and phase screens of dimension 100 by 100 are used for analysis in the experiments.

An optimum error threshold has to be chosen. Choosing a very tiny value will reduce the segment size and it may happen that the last segment size may not be sufficient enough to accurately predict the pixel variation trends. A large value of error threshold as expected leads to large errors. The choice of segmentation size and error threshold is thereby equivalent. The effect of choosing segment size is shown in the case of predicting turbulence phase screens for 4 realizations of time series data sets in Figure 5. On the whole, a segment size of 3 - 4 looks to be the optimum number.

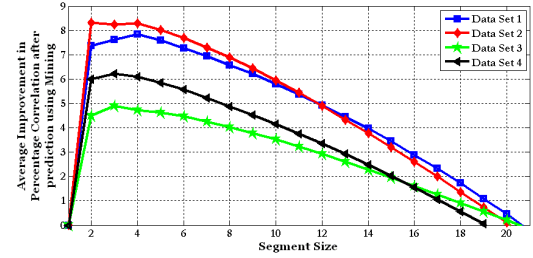

Fig. 5. Percentage improvement in prediction at different segment sizes

The ratio of the decorrelation time to servo lag error is the key factor that decides the phase screen that needs to be predicted. For example, if the decorrelation time is 15 milliseconds and the servo lag time is nearly 5 milliseconds, the ratio turns out to be 3. From Figure 3, the correlation goes below 25% (decorrelation cutoff) after 30 phase screens. This suggests that the phase screen that needs to be predicted is the tenth one which is correlated by 85% to the first one. Changing any of decorrelation time or servo lag error effectively changes these values. If prediction were not to be performed, we would be trying to compensate the wave-front using a wave-front that is only 85% correlated to it (reconstruction inefficiencies not included). This is the reason why the improvement due to prediction is calculated as the difference between the correlation of the last phase screen in the sub-data cube with the predicted one and the correlation of the last phase screen in the sub-data cube with the actual phase screen from the data cube. Decorrelation timescales are dependent on the seeing conditions and the servo lag error depends on the instrument speeds. The effect of varying ratio of decorrelation time to servo lag error is experimentally studied. The accuracy of predicting phase screens at different time lags within the decorrelation timescale is shown in Figure 6. In this figure, the y-axis is normalized with the amount of decorrelation between the last phase screen in the sub-data cube and the actual phase screen. If the decorrelation between the last phase screen in the sub-data cube and the actual phase screen is 0.1, then the percentage improvement in predicting the $10^{th}$ phase screen lies between 4.5-7.5% from the graph.

Prediction using linear interpolation, linear regression as well as higher order interpolation was performed. It was observed that interpolation using higher order polynomials generally fails in predicting the future phase screens accurately. This is because the choice of the polynomial fitting that must be used to pixel data sets is not consistent throughout the phase screen. Linear interpolation as well as linear regression methods performs better. Linear regression is although slower, the prediction performance on an average is better than linear interpolation by 1.5%.

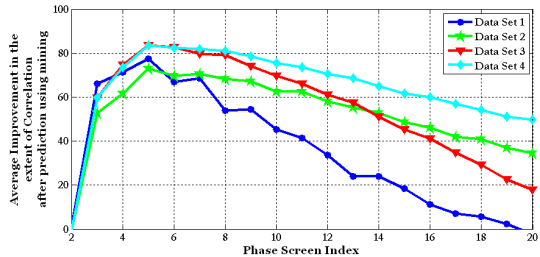

Fig. 6. Percentage improvement in prediction at different time lags

## 7. CONCLUSIONS

An attempt has been made to predict the incoming wave-fronts distorted due to atmospheric turbulence using data mining the time series data. This will help in improving the errors induced by the servo lag delays introduced by different components in adaptive optics imaging systems. On an average we report an improvement of 6% due to prediction using data mining. It is evident from the simulation analysis that the improvement depends on segmentation, decorrelation timescales, servo lag error timescales and the fitting methodology used. Very large and too small segmentation errors lead to worse prediction. Segmentation size of 3 or 4 was found to be optimum for best prediction. For the simulated phase screens, it was found that predicting the 5$^{th}$ phase screen gives maximum correlation, which corresponds to nearly 3 milliseconds servo lag error when the decorrelation time is 18 milliseconds. This study helps astronomers to optimize the exposure time in adaptive optical imaging systems.